\documentclass[10pt,a4paper,twocolumn,english,amssymb,nofootinbib,superscriptaddress,showpacs]{revtex4}
\usepackage{lmodern}

\usepackage[T1]{fontenc}
\usepackage[latin9]{inputenc}
\usepackage{babel}
\usepackage{float}
\usepackage{amsmath}
\usepackage{amssymb}
\usepackage{graphicx}
\usepackage{esint}
\usepackage[unicode=true,pdfusetitle,
 bookmarks=true,bookmarksnumbered=false,bookmarksopen=false,
 breaklinks=false,pdfborder={0 0 1},backref=false,colorlinks=false]
 {hyperref}

\makeatletter

\pdfpageheight\paperheight
\pdfpagewidth\paperwidth

\@ifundefined{textcolor}{}
{%
 \definecolor{BLACK}{gray}{0}
 \definecolor{WHITE}{gray}{1}
 \definecolor{RED}{rgb}{1,0,0}
 \definecolor{GREEN}{rgb}{0,1,0}
 \definecolor{BLUE}{rgb}{0,0,1}
 \definecolor{CYAN}{cmyk}{1,0,0,0}
 \definecolor{MAGENTA}{cmyk}{0,1,0,0}
 \definecolor{YELLOW}{cmyk}{0,0,1,0}
 }

\usepackage{latexsym}\usepackage{bm}

\makeatother

\begin{document}
\title{Spin-Spin Interactions in Massive Gravity and Higher Derivative Gravity Theories }

\author{\.{I}brahim G\"{u}ll\"{u} }

\email{ibrahimgullu2002@gmail.com}

\selectlanguage{english}%

\affiliation{Department of Physics,\\
 Middle East Technical University, 06800, Ankara, Turkey}

\author{Bayram Tekin}

\email{btekin@metu.edu.tr}

\selectlanguage{english}%

\affiliation{Department of Physics,\\
 Middle East Technical University, 06800, Ankara, Turkey}

\date{\today}
\begin{abstract}

We show that, in the weak field limit, at large separations, in sharp contrast to General Relativity (GR), all massive
gravity theories predict  distance-dependent spin alignments for spinning objects. For all separations GR requires 
anti-parallel spin orientations with spins pointing along the line joining the sources. Hence total spin is minimized in GR. On the other hand, while
massive gravity at small separations ($m_gr \le1.62$) gives the same result as GR, for large separations ($m_gr>1.62$)
the spins become parallel to each other and perpendicular to the line joining the objects.  
Namely, the  potential energy is minimized  when the total spin is maximized in  massive gravity for large separations. We also compute the spin-spin interactions in quadratic gravity theories and find that while at large separations GR result is intact, at small separations, spins become perpendicular to the line joining sources and anti-parallel to each other. 

\end{abstract}
\maketitle
{\bf Introduction:} 
Consider two widely separated spinning massive objects (for example
two galaxies or galaxy clusters) that interact via gravity: What is the
minimum energy configuration for their spin orientations, and how
does the result depend on whether the graviton is massive or not?
In this work we will compute the spin-spin interactions of point-like
objects in massive gravity. We will show that introducing  a small graviton mass gives the highly 
unexpected result of changing the spin orientations of sources from the one predicted in GR. 
Arguably, massive gravity is the most natural modification of  GR that has implications in the overall dynamics- accelerated 
expansion-of the universe and hence a detailed study of gravitomagnetic effects such as the one done in this work is needed.

Before we give a detailed derivation of the results in the next section in $D$ dimensional spacetimes and higher curvature theories, let us summarize our findings here for the case of $D=3+1$ for GR and massive gravity. Consider two localized spinning point-like sources described with the components of the energy momentum tensor
\begin{eqnarray} 
T_{00}&=&m_{a}\delta^{\left(3\right)}\left(\vec{x}-\vec{x}_{a}\right), \nonumber \\
T_{\phantom{i}0}^{i}&=&-\frac{1}{2}J_{a}^{k}\,\epsilon^{ikj}\partial_{j}\delta^{\left(3\right)}\left(\vec{x}
-\vec{x}_{a}\right), \label{en_mom}
\end{eqnarray}
where $a=1,\,2$. Here $m_{a}$ is the mass and $\vec{J_{a}}$ is
the spin of the particle. Then, working in a flat background, from
the tree-level diagram of one graviton exchange, we can calculate
the potential energy as 
\begin{equation}
U=-\frac{4\pi G}{t}\int d^{4}x\, d^{4}x^{\prime}T^{\mu\nu}\left(x\right)G_{\mu\nu\alpha\beta}
\left(x,x^{\prime}\right)T^{\alpha\beta}\left(x^{\prime}\right),\label{green}
\end{equation}
where $G_{\mu\nu\alpha\beta}\left(x,x^{\prime}\right)$ is the Green's
function of the theory at hand and $t$ is a large time that will
drop at the end. In GR this computation gives
\begin{equation}
U_{GR}=-\frac{Gm_{1}m_{2}}{r}-\frac{G}{r^{3}}\left[\vec{J_{1}}\centerdot\vec{J_{2}}-3\vec{J_{1}}\centerdot\hat{r}\,
\vec{J_{2}}\centerdot\hat{r}\right],\label{general_relativity}
\end{equation}
where $\vec{r}=r\hat{r}$ is the distance between the two sources.
Spin-spin part can be attractive or repulsive depending on the spin
orientations. Maximum value of $\vec{J_{1}}\centerdot\vec{J_{2}}-3\vec{J_{1}}\centerdot\hat{r}\,
\vec{J_{2}}\centerdot\hat{r}$,
that is the minimum of the potential energy is achieved when $\vec{J_{1}}$and
$\vec{J_{2}}$ are anti-parallel and point along $\hat{r}$ as depicted in Figure 1. That
means in GR, for \emph{any} \emph{given} $r$, potential energy is minimized
for anti-parallel spin orientations, if we neglect the tidal and orbital
angular momentum effects. (The computation here is of course not a good approximation for close binary systems, such as
two neutron stars {\it etc.}, but it is a valid approximation for two widely separated galaxies or galaxy clusters.)
\begin{figure}[h]
\includegraphics[width=\linewidth]{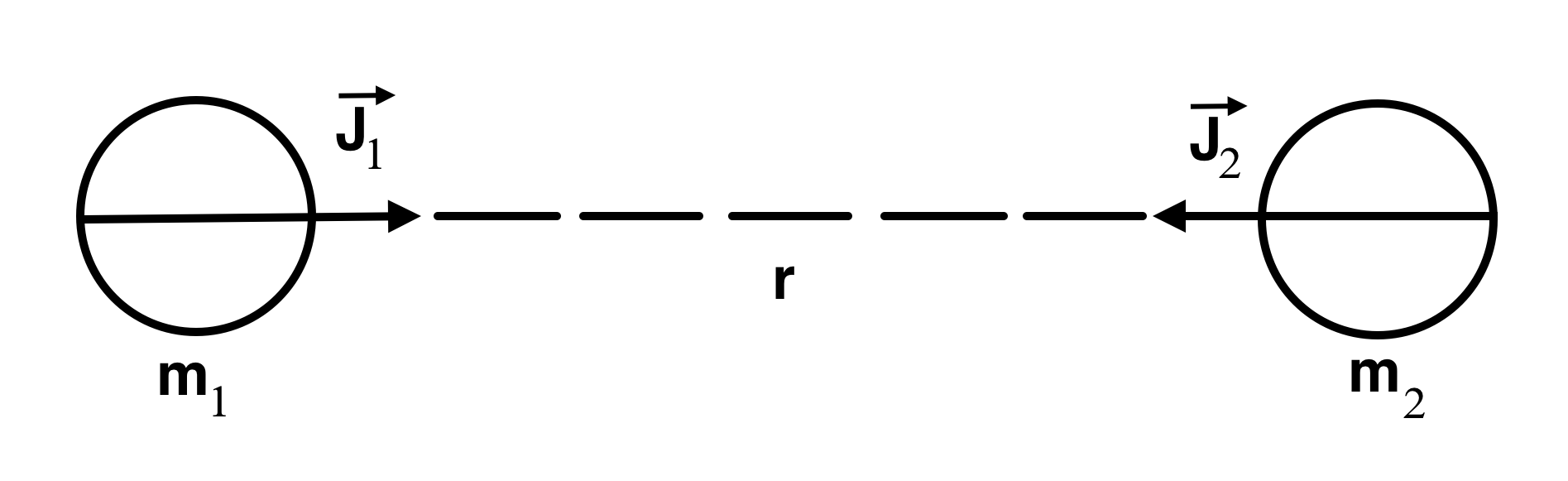}\caption{Minimum energy configuration in GR, as long as weak field limit is applicable.}
\end{figure}
Let us give the results of the same computation in massive gravity. At this point one might worry about which massive 
gravity to use. The crucial point is that in the  weak field limit around flat space, any viable (non-linear, ghost-free)
massive gravity theory reduces to the Fierz-Pauli (FP) theory that describes 5 degrees of freedom. Hence the following
computation is a universal, weak field, large distance, prediction of { \it all} massive gravity theories built to
describe 5 degrees of freedom around flat space.  The Lagrangian density of the  linear massive gravity is  
\begin{align}
\mathcal{L}_{FP} & =\frac{1}{16\pi G}\left[R-\frac{m_{g}^{2}}{4}\left(h_{\mu\nu}^{2}-h^{2}\right)\right]
+{\cal L}_{\text{matter}},\label{massive_lagrangian}
\end{align}
where $m_{g}$ is the mass of the graviton, we found that at the lowest order the potential energy is 
\begin{align}
U_{FP}= & -\frac{4}{3}Gm_{1}m_{2}\frac{e^{-m_{g}r}}{r}-\frac{Ge^{-m_{g}r}\left(1+m_{g}r
+m_{g}^{2}r^{2}\right)}{r^{3}}\label{massive_gravity}\\
 & \times\left[\vec{J_{1}}\centerdot\vec{J_{2}}-3\vec{J_{1}}\centerdot\hat{r}\,
 \vec{J_{2}}\centerdot\hat{r}\frac{\left(1+m_{g}r+\frac{1}{3}m_{g}^{2}r^{2}\right)}{\left(1+m_{g}r+m_{g}^{2}r^{2}\right)}
 \right].\nonumber
\end{align}
 It is clear that, in contrast to the GR result,  in massive gravity depending on the distance between
the sources, spin-spin part of the potential energy is minimized for different spin orientations determined 
by the maximization of  the function (see the Appendix for details) 
\begin{equation}
f\left(\theta,\varphi_{1},\varphi_{2}\right)=\cos\left(\theta\right)-3\frac{\left(1+x+\frac{1}{3}x^{2}\right)}
{\left(1+x+x^{2}\right)}\cos\left(\varphi_{1}\right)\cos\left(\varphi_{2}\right),\label{function}
\end{equation}
where $x=m_{g}r$ and $\theta$ is the angle between the spins and
$\varphi_{i}$ is the angle between $\vec{J_{i}}$ and $\vec{r}$.
Maximization of (\ref{function}) yields: anti-parallel spins for $x \le \frac{1+\sqrt{5}}{2}\approx1.62$ as in the case 
of GR depicted in Figure 2. On the other hand,  for $x>\frac{1+\sqrt{5}}{2}\approx1.62$, one gets parallel spins which are perpendicular 
to the line joining the 
sources as in Figure 3.
\begin{figure}[h]
\includegraphics[width=\linewidth]{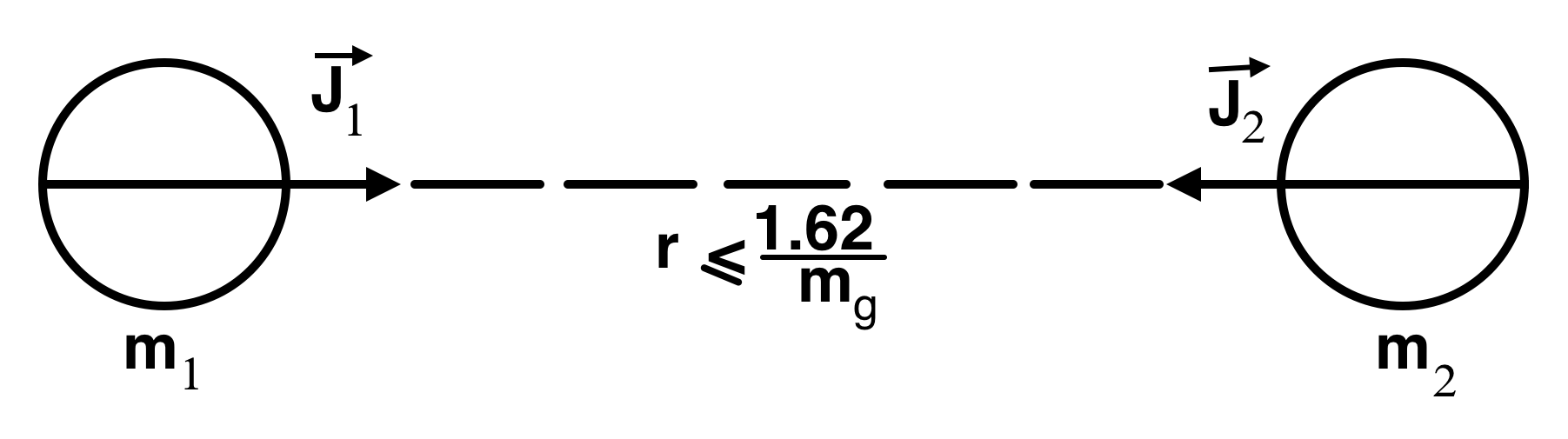}\caption{Minimum energy configuration in massive gravity for $m_gr\leq 1.62$.}
\end{figure}
\begin{figure}[h]
\includegraphics[width=\linewidth]{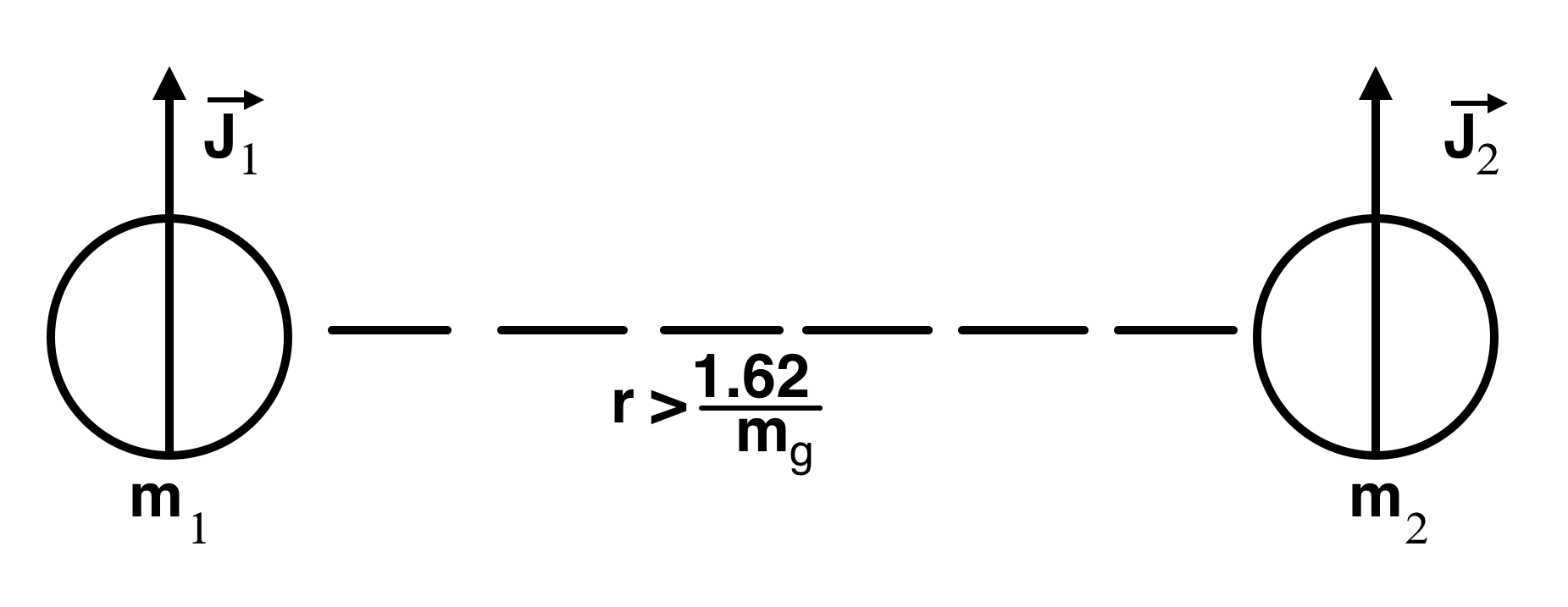}\caption{In massive gravity, at large separations, 
the potential energy is minimized when the spins
are perpendicular to the line joining the sources.}
\end{figure}
The important conclusion one learns is that while in GR  minimal potential energy is realized for minimum total spin at all separations,
in massive gravity potential energy is minimized for maximum total spin for $ m_g r > 1.62$.~\footnote{We would like to thank A. Dane whose simulation of the spin-spin interaction led us to realize this point where spins suddenly change orientations. Note that the same  point  that is the "Golden Number" arises when one considers stable circular orbits in the Newtonian theory with a Yukawa potential. Namely, stable circular orbits exist for  $x \le  \frac{1+\sqrt{5}}{2}$. We thank F. \"{O}ktem for this point. }

{\bf Derivation of the results: } 
To derive the above results and their $D$ dimensional generalizations
in GR, massive gravity and quadratic gravity, it is somewhat more convenient
to use the propagator found in \cite{GulluTekin} to represent (\ref{green}). In order to avoid repeating the computations of all three theories
 let us consider the most general theory which includes these theories: 
\begin{eqnarray}
S & = & \int d^{D}x\,\sqrt{-g}\left\{ \frac{1}{\kappa}R-\frac{2\Lambda_{0}}{\kappa}+\alpha R^{2}+\beta R_{\mu\nu}^{^{2}} \right. \nonumber \\ 
&& \left.+\gamma\left(R_{\mu\nu\sigma\rho}^{2}-4R_{\mu\nu}^{2}+R^{2}\right)\right\} \nonumber \\
 &  & +\int d^{D}x\,\sqrt{-g}\left\{ -\frac{m_{g}^{2}}{4\kappa}\left(h_{\mu\nu}^{2}-h^{2}\right)+{\cal L}_{\text{matter}}\right\} ,\label{action}
\end{eqnarray}
 In \cite{GulluTekin}, we computed
the scattering amplitude ( $A=Ut$) corresponding to a graviton exchange
in this theory and presented it with sufficient detail, hence we quote
here the result: 
\begin{eqnarray}
4A & = & 2 T_{\mu\nu}^{\prime}\left\{ (\beta\bar{\square}+a)(\triangle_{L}^{(2)}-\frac{4\Lambda}{D-2})+\frac{m_{g}^{2}}{\kappa}\right\} ^{-1}T^{\mu\nu} \nonumber \\
 & + & \frac{2}{D-1}T^{\prime}\left\{ (\beta\bar{\square}+a)(\bar{\square}+\frac{4\Lambda}{D-2})-\frac{m_{g}^{2}}{\kappa}\right\} ^{-1}T
 \nonumber \\
 &-&  \frac{4\Lambda}{(D-2)(D-1)^{2}}T^{\prime}\left\{ (\beta\bar{\square}+a)(\bar{\square}+\frac{4\Lambda}{D-2})-\frac{m_{g}^{2}}{\kappa}\right\} ^{-1}  \nonumber \\
 &&\times \left\{ \bar{\square}+\frac{2\Lambda D}{(D-2)(D-1)}\right\} ^{-1}T\nonumber \\
 & + & \frac{2}{(D-2)(D-1)}T^{\prime}\left\{ \frac{1}{\kappa}+4\Lambda f-c\bar{\square}-\frac{m_{g}^{2}}{2\kappa\Lambda}(D-1)\right\} ^{-1} \nonumber \\
&&\times \left\{ \bar{\square}+\frac{2\Lambda D}{(D-2)(D-1)}\right\} ^{-1}T, \label{mainresult}
\end{eqnarray}
where we have dropped the integral signs not to clutter the notation and also to properly account all those theories in the corresponding limits,
we have provisionally introduced an effective cosmological constant
which is determined via the quadratic equation $\frac{\Lambda-\Lambda_{0}}{2\kappa}+f\Lambda^{2}=0.$
The other parameters that appear above are defined as 
\begin{eqnarray}
 & f & \equiv\left(D\alpha+\beta\right)\frac{\left(D-4\right)}{\left(D-2\right)^{2}}+\gamma\frac{\left(D-3\right)\left(D-4\right)}{\left(D-1\right)\left(D-2\right)},\\
 & a & \equiv\frac{1}{\kappa}+\frac{4\Lambda D}{D-2}\alpha+\frac{4\Lambda}{D-1}\beta+\frac{4\Lambda\left(D-3\right)\left(D-4\right)}{\left(D-1\right)\left(D-2\right)}\gamma,\\
 & c & =\frac{4(D-1)\alpha+D\beta}{D-2}.
\end{eqnarray}
 With all these parameters at hand, one covers all the three theories
that we are interested in. For example the result for General Relativity
follows from $m_{g}^{2}=\alpha=\beta=\gamma=0$ which yield $a=\frac{1}{\kappa}$
and $f=c=0$. For flat backgrounds one has 
\begin{equation}
4A=-2\kappa T{}_{\mu\nu}^{\prime}(\partial^{2})^{-1}T^{\mu\nu}+\frac{2\kappa}{D-2}T^{\prime}(\partial^{2})^{-1}T.
\label{gr_amp}
\end{equation}
More explicitly the last equation  is 
\begin{eqnarray}
&4A&= -2\kappa\int d^{D}x\int d^{D}x^{\prime}T_{\mu\nu}\left(x^{\prime}\right)G\left(x,x^{\prime}\right)T^{\mu\nu}\left(x\right) \nonumber \\
&+&\frac{2\kappa}{\left(D-2\right)}\int d^{D}x\int d^{D}x^{\prime}T\left(x^{\prime}\right)G\left(x,x^{\prime}\right)T\left(x\right),
\end{eqnarray}
 where the scalar Green's function reads 
\[
\partial_{x}^{2}G\left(x,x^{\prime}\right)=-\delta^{D}\left(x,x^{\prime}\right),
\]
 and $\partial_{x}^{2}=-\partial_{t}^{2}+\vec{\nabla}^{2}$. Of course
one must keep in mind that to reach the explicit final result for
the potential energies one uses 

\begin{eqnarray}
\left(\partial^{2}\right)^{-1}\equiv G_{R}\left(x,\, x^{\prime}\right)=\frac{\Gamma\left(\frac{D-3}{2}\right)}{4\pi^{\frac{D-1}{2}}r^{D-3}}\delta\left[r-\left(t-t^{\prime}\right)\right],\label{laplacian_green}
\end{eqnarray}
 in the massless case and similarly for the massive case 
\begin{eqnarray}
 G_{R}\left(x,\, x^{\prime}\right) 
=\frac{\left(\frac{m_{g}}{r}\right)^{\frac{D-3}{2}}}{\left(2\pi\right)^{\frac{D-1}{2}}}K_{\frac{D-3}{2}}\left(r\, m_{g}\right)\delta\left[r-\left(t-t^{\prime}\right)\right],\label{massive_greens}
\end{eqnarray}
 for the retarded Green's functions.

We are now ready to compute the potential energy for the desired theory.
We will give two explicit examples below: GR ( Einstein's theory )
and Fierz-Pauli massive gravity. The analogous computations in the
quadratic theory, without an explicit mass term follow similarly.

For massive spinning sources the energy-momentum tensor is given as the $D$-dimensional generalization of (\ref{en_mom}) 
\begin{eqnarray}
&&T_{00}=m_{a}\delta^{\left(D-1\right)}\left(\vec{x}-\vec{x}_{a}\right), \, \,\,\, T_{ij}=0, \nonumber \\
&& T_{\phantom{i}0}^{i}=-\frac{1}{2}J^{klm...}\,\epsilon^{iklm...j}\partial_{j}\delta^{\left(D-1\right)}\left(\vec{x}-\vec{x}_{a}\right ),\label{Dspin}
\end{eqnarray}
 where $J$ has $D-3$ and $\epsilon$ has $D-1$ indicies and $a=1,2$.  Generators
of rotations will be given as $M^{ij}=\int d^{D-1}x\;\left(x^{i}T_{\phantom{j}0}^{j}-x^{j}T_{\phantom{i}0}^{i}\right)$,
which yields $M^{ij}=\epsilon^{ijk}J^{k}$ in $D=3+1$. It is important to note that taking the background to be the Minkowski
space, with $\eta_{\mu\nu}=\mbox{diag}(-,+,...+)$, one has $T=\eta^{\mu\nu}T_{\mu\nu}=-T_{00}$
and hence the trace part does not play a role in the spin-spin interactions. 

{\bf Spin-spin interaction in General Relativity:}
For massless gravity in $D$-dimensions, we have 
\begin{eqnarray}
4A 
  =&-&2\kappa T{}_{00}^{\prime}\left\{ \frac{1}{\partial^{2}}\right\} T^{00}
+\frac{2\kappa}{(D-2)}T^{\prime}\left\{ \frac{1}{\partial^{2}}\right\} T  \nonumber \\
&-&4\kappa T{}_{0i}^{\prime}\left\{ \frac{1}{\partial^{2}}\right\} T^{0i}.\label{amp_d}
\label{massless_amp}
\end{eqnarray}
 The first two terms give $4t$ times the usual  Newtonian potential energy which we need not depict here. The last term, which is the relevant part for spin-spin interactions, reads 
\begin{eqnarray}
&&-4\kappa T{}_{0i}^{\prime}\left\{ \frac{1}{\partial^{2}}\right\} T^{0i} = \nonumber \\
 && 2\kappa J_{1}^{a_{1}a_{2}...a_{D-3}}\,\epsilon^{ia_{1}a_{2}...a_{D-3}n}\partial_{n}^{\prime}\delta^{\left(D-1\right)}\left(\vec{x}^{\prime}-\vec{x}_{1}\right) \nonumber \\
 && \times\left\{ \frac{1}{\partial^{2}}\right\} \left(\frac{1}{2}\right)J_{2}^{b_{1}b_{2}...b_{D-3}}\,\epsilon^{ib_{1}b_{2}...b_{D-3}m}\partial_{m}\delta^{\left(D-1\right)}\left(\vec{x}-\vec{x}_{2}\right) \nonumber \\
= && \kappa J_{1}^{a_{1}a_{2}...a_{D-3}}J_{2}^{b_{1}b_{2}...b_{D-3}}\,\epsilon^{ia_{1}a_{2}...a_{D-3}n}\epsilon^{ib_{1}b_{2}...b_{D-3}m}\nonumber \\
 && \times\partial_{n}^{\prime}\delta^{\left(D-1\right)}\left(\vec{x}^{\prime}-\vec{x}_{1}\right)\left\{ \frac{1}{\partial^{2}}\right\} \partial_{m}\delta^{\left(D-1\right)}\left(\vec{x}-\vec{x}_{2}\right).
\end{eqnarray}
 This expression looks somewhat cumbersome, to understand the crux
of the computation, let us carry it  out  more explicitly
in $D=3+1$ dimensions. 
\begin{eqnarray}
&&-4\kappa T{}_{0i}^{\prime}\left\{ \frac{1}{\partial^{2}}\right\} T^{0i}= \nonumber \\
&&4\kappa\int  d^{4}x'\int d^{4}x\frac{1}{4}J_{1}^{k}\epsilon^{ikj}\partial_{j}\delta^{\left(3\right)}\left(\vec{x}-\vec{x}_{1}\right) \\
 && \times\frac{1}{4\pi|\vec{x}-\vec{x'}|}\delta\big(|\vec{x}-\vec{x'}|-(t-t')\big)J_{2}^{l}\epsilon^{ilm}\partial'_{m}\delta^{\left(3\right)}\left(\vec{x'}-\vec{x}_{2}\right).\nonumber
\end{eqnarray}
 Carrying out the time integrals and performing integration by parts,
one gets 
\begin{eqnarray}
&&-4\kappa T{}_{0i}^{\prime}\left\{ \frac{1}{\partial^{2}}\right\} T^{0i}  = \nonumber \\
 && \frac{t\kappa}{4\pi}\Big(\delta^{ij}\vec{J_{1}}\cdot\vec{J_{2}}-J_{1}^{i}J_{2}^{j}\Big)\frac{\partial}{\partial x_{1}^{i}}\frac{\partial}{\partial x_{2}^{j}}\frac{1}{|\vec{x}_{1}-\vec{x}_{2}|}.
\end{eqnarray}
 Since the sources do not coincide, $\vec{x}_{1}\ne\vec{x}_{2}$, one has 
\begin{equation}
\frac{\partial}{\partial x_{1}^{i}}\frac{\partial}{\partial x_{2}^{j}}\frac{1}{|\vec{x}_{1}-\vec{x}_{2}|}=\frac{1}{r^{3}}\Big(\delta^{ij}-3\hat{r}^{i}\hat{r}^{j}\Big),
\end{equation}
and therefore spin-spin interaction potential energy of GR is found (\ref{general_relativity}). 

In $D$ dimensions contractions
of the $\epsilon$ tensor only change the relative coefficients of the
two terms in the spin-spin part. To obtain the generic result, one
way is to do the computation in  several other dimensions and find the formula or
one can use the contractions of the $\epsilon$ tensor. We have done both
ways, the result is 
\begin{eqnarray}
&&-4\kappa T{}_{0i}^{\prime}\left\{ \frac{1}{\partial^{2}}\right\} T^{0i}= \nonumber \\
 && \left(D-3\right)!\kappa\Big\{J_{1}^{a_{1}a_{2}...a_{D-3}}J_{2}^{a_{1}a_{2}...a_{D-3}}\partial_{m}\partial_{m}^{\prime}\nonumber \\
&&-\left(D-3\right)J_{1}^{a_{1}a_{2}...a_{D-4}m}J_{2}^{a_{1}a_{2}...a_{D-4}n}\partial_{m}\partial_{n}^{\prime}\Big\} \nonumber \\
 && \times\left\{ \frac{1}{\partial^{2}}\right\} \delta^{\left(D-1\right)}\left(\vec{x}^{\prime}-\vec{x}_{1}\right)\delta^{\left(D-1\right)}\left(\vec{x}-\vec{x}_{2}\right).
\end{eqnarray}
 Finally the spin-spin interaction in $D$-dimensional massless gravity reads  
\begin{eqnarray}
U_{GR}= &&  -\frac{G_{D}\left(D-2\right)!\left(D-3\right)^{2}}{2r^{D-1}}\nonumber \\
&&\times\Big(J_{1}\centerdot J_{2}-\left(D-1\right)\left(J_{1}\centerdot\hat{r}\right)\left(J_{2}\centerdot \hat{r}\right)\Big),
\end{eqnarray}
 where the $D$-dimensional Newton's constant is  
\[
G_{D}=\frac{\kappa\Gamma\left(\frac{D-3}{2}\right)}{8\left(D-2\right)\pi^{\frac{D-1}{2}}},
\]
 which gives $\kappa=16$$\pi G$ in $D=3+1$. Here we defined the scalar products between the anti-symmetric objects as 
\begin{eqnarray}
J_{1}\centerdot J_{2}&\equiv& J^{a_{1}a_{2}...a_{D-3}}J^{a_{1}a_{2}...a_{D-3}} \\
\left(J_{1}\centerdot\hat{r}\right)\left(J_{2}\centerdot\hat{r}\right)&\equiv& J^{a_{1}a_{2}...a_{D-3}}\hat{r}_{D-3}J^{a_{1}a_{2}...a_{D-3}}\hat{r}_{D-3}\nonumber.
\end{eqnarray}

{\bf Scattering in Massive $D$-Dimensional Gravity: }
Let us now do the same computation in the linearized massive gravity. The relevant scattering
amplitude is

\begin{eqnarray}
4A&=& -2\kappa T{}_{00}^{\prime}\left\{ \partial^{2}-m_{g}^{2}\right\} ^{-1}T^{00} \nonumber \\
&+&\frac{2\kappa}{\left(D-1\right)}T^{\prime}\left\{ \partial^{2}-m_{g}^{2}\right\} ^{-1}T \nonumber \\
&-&4\kappa T{}_{0i}^{\prime}\left\{ \partial^{2}-m_{g}^{2}\right\} ^{-1}T^{0i} \nonumber \\
&=& -2\kappa\left(\frac{D-2}{D-1}\right)m_{1}m_{2}\frac{1}{\left(2\pi\right)^{\frac{D-1}{2}}}\frac{1}{r^{\frac{D-3}{2}}} \nonumber \\
&\times&\left[\left(\frac{1}{m_{g}^{2}}\right)^{\frac{3-D}{4}}K_{\frac{D-3}{2}}\left(r\, m_{g}\right)\right] \nonumber \\
 &+&\kappa\left(D-3\right)!\Big[J_{1}^{a_{1}a_{2}...a_{D-3}}J_{2}^{a_{1}a_{2}...a_{D-3}}\partial_{m}\partial_{m}^{\prime} \nonumber \\
&-& \left(D-3\right)J^{a_{1}a_{2}...a_{D-4}m}J^{a_{1}a_{2}...a_{D-4}n}\partial_{m}\partial_{n}^{\prime}\Big]\nonumber \\
 && \times\frac{1}{\left(2\pi\right)^{\frac{D-1}{2}}}\frac{1}{r^{\frac{D-3}{2}}}\left[\left(\frac{1}{m_{g}^{2}}\right)^{\frac{3-D}{4}}K_{\frac{D-3}{2}}\left(r\, m_{g}\right)\right].
\end{eqnarray}
 Just like the massless case that we studied in detail in the previous
section, one performs partial integrations and carries out the integrals to get 
\begin{align}
U_{FP}= & -\frac{\kappa\left(D-2\right)m_{1}m_{2}\; m_{g}^{\frac{D-3}{2}}}{2\left(D-1\right)\left(2\pi\right)^{\frac{D-1}{2}}
r{}^{\frac{D-3}{2}}}
K_{\frac{D-3}{2}}\left(r\, m_{g}\right)\nonumber \\
 & +\frac{\kappa\left(D-3\right)!\, m_{g}^{\frac{D+1}{2}}}{4\left(2\pi\right)^{\frac{D-1}{2}}r{}^{\frac{D-3}{2}}}K_{\frac{D+1}{2}}
 \left(r\, m_{g}\right)
 \nonumber \\
 & \times\left[J_{1}\centerdot J_{2}\left(\frac{2K_{\frac{D-1}{2}}\left(r\, m_{g}\right)}{r\, m_{g}K_{\frac{D+1}{2}}\left(r\,
 m_{g}\right)}-1\right)\right.\nonumber \\
 & \left.\phantom{\frac{A_{D}}{B_{D}}}+\left(D-3\right)\left(J_{1}\centerdot\hat{r}\right)\left(J_{2}\centerdot\hat{r}\right)\right]
 .\label{d_massive_potential}
\end{align}
For $D=3+1$ (\ref{d_massive_potential}) gives (\ref{massive_gravity}).

{\bf Quadratic Gravity:} With the tools at our hand we can extend the above results to $D$
dimensional quadratic gravity without a Fierz-Pauli mass term with the Lagrangian
density 
\begin{eqnarray*}
\mathcal{L} & = & \frac{1}{\kappa}R+\alpha R^{2}+\beta R_{\mu\nu}^{^{2}}+\gamma\left(R_{\mu\nu\sigma\rho}^{2}
-4R_{\mu\nu}^{2}+R^{2}\right).
\end{eqnarray*}
The amplitude can be written as 
\begin{align}
U_{quad}\times t & =-\frac{\kappa}{2}T{}_{\mu\nu}^{\prime}\left(\partial^{2}\right)^{-1}T^{\mu\nu}+\frac{\kappa T^{\prime}
\left(\partial^{2}\right)^{-1}T}{2\left(D-2\right)}\nonumber \\
 & +\frac{\kappa}{2}T{}_{\mu\nu}^{\prime}\left(\partial^{2}-m_{\beta}^{2}\right)^{-1}T^{\mu\nu}-\frac{\kappa T^{\prime}
 \left(\partial^{2}-m_{\beta}^{2}\right)^{-1}T}{2\left(D-1\right)}\nonumber \\
 & -\frac{\kappa T^{\prime}\left(\partial^{2}-m_{c}^{2}\right)^{-1}T}{2\left(D-2\right)\left(D-1\right)},\label{quad_amplitude}
\end{align}
where $m_{\beta}^{2}=-\frac{1}{\kappa\beta}$ and $m_{c}^{2}=\frac{1}{\kappa\left(4\alpha\left(D-1\right)+D\beta\right)}$.
All the terms in the above expression have been computed above: The
first line is pure GR, the second and third lines come from the quadratic
terms in the Lagrangian. The fourth and the fifth terms do not contribute
to the spin-spin interactions, the third term gives a negative contribution
to the spin-spin interaction in comparison with the GR result. The
full expression is somewhat cumbersome to depict, $D=2+1$ case was
given in \cite{Canonic}, here let us write down the D = 3 +1 result. 

Let $U_{quad}\equiv U_{GR}+U_{2}$, then the contribution coming from
the quadratic part reads 
\begin{align}
U_{2}= & \frac{Gm_{1}m_{2}}{r}\left(\frac{4}{3}e^{-m_{\beta}r}-\frac{1}{3}e^{-m_{c}r}\right)\nonumber \\
 & +\frac{Ge^{-m_{\beta}r}\left(1+m_{\beta}r+m_{\beta}^{2}r^{2}\right)}{r^{3}}\nonumber \\
 & \times\left[\vec{J_{1}}\centerdot\vec{J_{2}}-3\vec{J_{1}}\centerdot\hat{r}\,\vec{J_{2}}\centerdot\hat{r}
 \frac{\left(1+m_{\beta}r+\frac{1}{3}m_{\beta}^{2}r^{2}\right)}{\left(1+m_{\beta}r+m_{\beta}^{2}r^{2}\right)}\right].
 \label{quad_corrections}
\end{align}
At long distances, GR part dominates and hence spins are anti-parallel
to each other and point along $\hat{r}$. In short distances quadratic
part dominates and spin-spin interaction part is just like the one
in massive gravity but with an overall negative sign. Therefore, for
quadratic gravity, at short distances spins are anti-parallel to each
other but they are perpendicular to $\hat{r}$ as shown in Fig 4.
\begin{figure}[h]
\includegraphics[width=\linewidth]{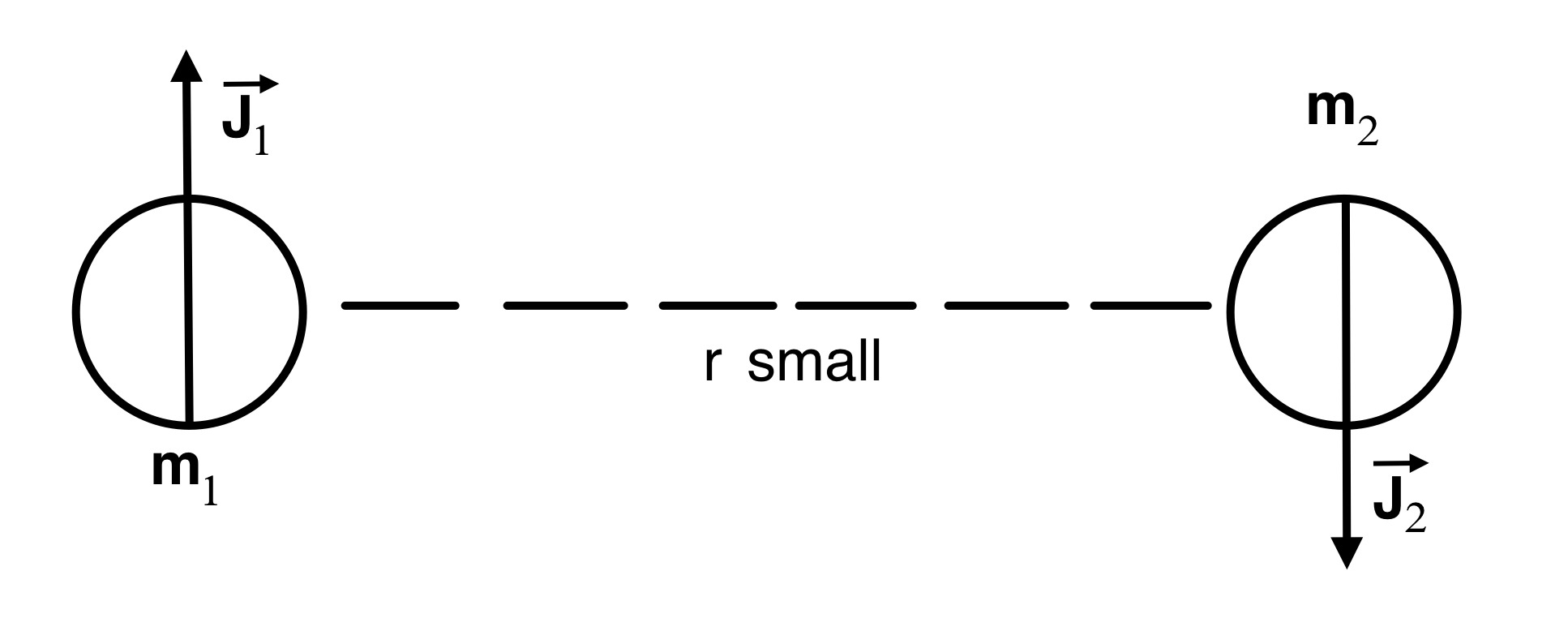}\caption{Minimum energy configuration in quadratic gravity for small separations.}
\end{figure}

{\bf  Conclusions and Discussions:}
 In this work, we have initiated a study of (linear) gravitomagnetic effects in the Fierz-Pauli 
massive gravity which is the unique linearized massive spin-2 theory describing 5 degrees of freedom in 3+1 dimensional 
flat backgrounds. ( Non-linear extensions of  the FP theory such as the dGRT theory \cite{dgrt} or its extensions which are free of 
the Boulware-Deser ghost \cite{bd}, though they still could be acausal \cite{dw}, yield exactly the same prediction as the FP 
theory at large distances in the weak field limit.) 
 
For two point-like spinning sources that interact gravitationally, potential energy is minimized for anti-parallel spin 
orientations pointing in the direction of the vector between the sources in General Relativity at any distance where the 
linear approximation is valid. On the other hand for massive gravity, potential energy is minimized  when the spins point 
away from the line joining the sources at large separations. Hence the total spin of the system is non-zero even for equal magnitude spins and  
point perpendicular to the axis joining the sources. 

A word about the mass-mass term in (\ref{massive_gravity}) is needed:  At large distances it is in the desired Yukawa like form 
which is one of the main motivations of studying massive gravity theories, since it can replace all or part  of the  dark energy 
needed to explain the accelerated expansion  of the Universe. As is clear that term also has the undesired  vDVZ discontinuity 
\cite{DV,Zak} in the vanishing graviton mass limit. Therefore  (\ref{massive_gravity}) cannot be applied to the scales where 
Newtonian ( or Einsteinian) gravity is well-tested. For such scales non-linear effects, such as the Vainshtein mechanism (\cite{Vain} 
or non-linear extensions of massive gravity, such as \cite{dgrt}, should come into play to correctly reproduce the observations 
within massive gravity.  The relevant scales that appear depend on the specific non-linear extension of massive gravity.  
See {\cite{hinterbichler} for an extensive review of massive gravity theories and how Vainshtein radius, below which non-linear 
theories must be used, can be set to the size of the solar system or the size of the galaxy. The point of view in 
this current work is that at sufficiently large separations where massive gravity is expected to deviate from GR, 
(\ref{massive_gravity}) describes the lowest order potential energy for all viable massive gravity theories  that reduce to 
the Fierz-Pauli theory at the linear level. [Of course, there may not exist a viable non-linear massive gravity theory free of the
ghost, acausality, strong coupling and vDVZ problems, but at this stage, there is still hope.] 

One could argue that compared to the Newtonian potential energy between the sources, the spin-spin potential energy is rather 
small and does not contribute much to the overall force. While this is correct, the overall force is not the relevant issue here: 
spin-spin force is quite distinct from the mass-mass force. The former is the sole force that determines the spin orientations.The 
situation is similar to the magnetic force in electrodynamics: While the magnetic force between two slowly moving charges with 
magnetic dipole moments (spins) is much smaller compared to the Coulomb force, it has a distinct effect on the charges.  In fact 
interacting  magnetic-dipole moments of charged particles give rise to ferromagnetic effects. In the context of massive gravity, 
a similar situation arises: spin-spin interaction of galaxies give rise to an overall spin of the system. Of course to derive observable consequences 
from our calculations above, one must carry out an $N$-body simulation of  galaxies.  The situation is actually quite similar to the Heisenberg model 
of three dimensional spins. It is an open question to see if  massive gravity could explain the observations of \cite{longo,shamir} who found that galaxies in a region have a non-zero total spin which cannot be easily explained by GR. 

This work is supported by the T\"{U}B\.{I}TAK Grant 113F155. We would like to thank T. \c{C}. \c{S}i\c{s}man, A. Karasu , S. Deser, A. Dane and F. \"{O}ktem for useful discussions.

{\bf APPENDIX: Finding the spin-orientations in GR and in massive gravity }
Here let us derive the minimum energy configuration for the spins
in both GR and massive gravity. The relevant part to be maximized
in the potential energy is 
\begin{equation}
h\equiv\vec{J}_{1}\cdot\vec{J}_{2}-f\left(x\right)\,\vec{J}_{1}\cdot\hat{r}\,\vec{J}_{2}\cdot\hat{r},\label{h_}
\end{equation}
where $x=m_{g}r$ and $f\left(x\right)=3$ for GR and the general
form of it is 
\begin{equation}
f\left(x\right)=\frac{3\left(1+x+\frac{1}{3}x^{2}\right)}{\left(1+x+x^{2}\right)}.\label{f_}
\end{equation}
Note that for massive gravity $f\left(x\right)\in\left[3,\,1\right)$ 

In spherical coordinates let us choose the plane of $\vec{J}_{1}$
and $\hat{r}$ as the $xy$-plane, and choose the direction of $\hat{r}$
as the $x$-axis. Therefore, $\vec{J}_{1}$ and $\vec{J}_{2}$ have
the following components in this coordinate system 
\begin{equation}
\vec{J}_{1}=J_{1}\left(\cos\varphi_{1}\hat{i}+\sin\varphi_{1}\hat{j}\right),\label{J_1}
\end{equation}
and 
\begin{equation}
\vec{J}_{2}=J_{2}\left(\cos\varphi_{2}\sin\theta_{2}\hat{i}+\sin\varphi_{2}\sin\theta_{2}\hat{j}+\cos\theta_{2}\hat{k}\right).\label{J_2}
\end{equation}
 Then, the relevant scalar products read 
\begin{equation}
\vec{J}_{1}\cdot\hat{r}=J_{1}\cos\varphi_{1}, \,\,\,\, \vec{J}_{2}\cdot\hat{r}=J_{2}\cos\varphi_{2}\sin\theta_{2},\label{J_2_dot_r} 
\end{equation} 
\begin{equation}
\vec{J}_{1}\cdot\vec{J}_{2}=J_{1}J_{2}\left(\cos\varphi_{1}\cos\varphi_{2}\sin\theta_{2}+\sin\varphi_{1}\sin\varphi_{2}\sin\theta_{2}\right).
\label{J_1_dot_J_2}
\end{equation}
Then (\ref{h_}) becomes 
\begin{align}
h= & J_{1}J_{2}\left[\cos\varphi_{1}\cos\varphi_{2}\sin\theta_{2}\left(1-f\right)+\sin\varphi_{1}\sin\varphi_{2}\sin\theta_{2}\right],\label{h(J_1,J_2)}
\end{align}
where we wrote $f\left(x\right)=f$. From (\ref{h(J_1,J_2)}) we see
that $\vec{J}_{1}$ and $\vec{J}_{2}$ must be on the same plane,which
follows from $\frac{\partial h}{\partial\theta_{2}}=0,\,\theta_{2}=\pm\frac{\pi}{2}$.
When these are put into (\ref{h(J_1,J_2)}) we see that $h$ becomes
a maximum for $\theta_{2}=\frac{\pi}{2}$ and a minimum for $\theta_{2}=-\frac{\pi}{2}$.
Since we want it to be a maximum (to get the minimum of the potential
energy) we choose $\frac{\pi}{2}$. Then 
\begin{align}
h & =J_{1}J_{2}\left[\cos\varphi_{1}\cos\varphi_{2}\left(1-f\right)+\sin\varphi_{1}\sin\varphi_{2}\right],\label{h_max_1}
\end{align}
and extremization with respect to two angles yield
\begin{align}
\frac{\partial h}{\partial\varphi_{1}} & =-\text{sin}\varphi_{1}\cos\varphi_{2}\left(1-f\right)+\text{cos}\varphi_{1}\sin\varphi_{2}=0,
\label{del_h_over_del_phi_1}\\
\frac{\partial h}{\partial\varphi_{2}} & =-\cos\varphi_{1}\text{sin}\varphi_{2}\left(1-f\right)+\sin\varphi_{1}\text{cos}\varphi_{2}=0,
\label{del_h_over_del_phi_2}
\end{align}
From now on the discussion bifurcates whether $f$ is $1$ or not. 

Let us first take $f=1$ then (\ref{del_h_over_del_phi_1}) and (\ref{del_h_over_del_phi_2})
become
\begin{align}
\text{cos}\varphi_{1}\sin\varphi_{2} & =0,\label{coupled_1}\\
\sin\varphi_{1}\text{cos}\varphi_{2} & =0.\label{coupled_2}
\end{align}
From (\ref{coupled_1}) $\varphi_{1}=\frac{\pi}{2}$ or $\varphi_{2}=0$
and from (\ref{coupled_2}) $\varphi_{1}=0$ or $\varphi_{2}=\frac{\pi}{2}$.
Therefore we have two solutions that are 
\begin{align}
\varphi_{1}=\varphi_{2} & =0,\nonumber \\
\varphi_{1}=\varphi_{2} & =\frac{\pi}{2}.\label{two_solutions}
\end{align}
Putting (\ref{two_solutions}) into (\ref{h(J_1,J_2)}) we get 
\begin{align}
h\left(\varphi_{1}=0,\,\varphi_{2}=0\right) & =0,\label{h_for_f_1_1}\\
h\left(\varphi_{1}=\frac{\pi}{2},\,\varphi_{2}=\frac{\pi}{2}\right) & =J_{1}J_{2},\label{h_for_f_1_2}
\end{align}
where (\ref{h_for_f_1_2}) gives the minimum potential energy. Both
spins point in the same direction and they are perpendicular to $\vec{r}$
joining the sources.

Let us continue our discussion with $f\neq1$: We plug (\ref{del_h_over_del_phi_1})
into (\ref{del_h_over_del_phi_2}) to get 
\begin{align}
\left[\left(1-f\right)^{2}-1\right]\text{sin}\varphi_{1}\cos\varphi_{2} & =0.\label{del_h_1_into_del_h_2}
\end{align}
There are again two cases which must be analyzed separately. One is
\begin{align}
\left(1-f\right)^{2}-1=0 & \Rightarrow f\left(f-2\right)=0.\label{f_cases}
\end{align}
For this case $f$ can be either $0$ or $2$. We know that $f$ is
in between $\left[3,1\right)$. Then $f$ cannot be $0$. Therefore,
it must be $2$. If $f\neq2$ then $\text{sin}\varphi_{1}\cos\varphi_{2}=0$
which is the second case. Before going into the details of the second
option, let us exhaust the first one:
\begin{align}
f=2\Rightarrow & \frac{3\left(1+x+\frac{1}{3}x^{2}\right)}{\left(1+x+x^{2}\right)}=2,\nonumber \\
x^{2}-x-1 & =0,\label{quadratic_x}
\end{align}
whose physical solution is
\[
x=\frac{1+\sqrt{5}}{2}\cong1.62.
\]
Note that at this point,
\begin{equation}
h=-J_{1}J_{2}\text{cos}\left(\varphi_{1}+\varphi_{2}\right),\label{h_critique}
\end{equation}
which is maximized for $\varphi_{1}+\varphi_{2} =\pi$, that is the same as the 
GR case. Lets look at the $f\neq2$ case. For this case 
\[
\text{sin}\varphi_{1}\cos\varphi_{2}=0.
\]
Then we have two possibilities that are $\varphi_{1}=0$ or $\pi$
and $\varphi_{2}$ is arbitrary or $\varphi_{2}=\frac{\pi}{2}$ or
$\frac{3\pi}{2}$ and $\varphi_{1}$ is arbitrary. Put $\varphi_{1}=0$
or $\varphi_{1}=\pi$ (both will give the same result) into (\ref{h_max_1})
\begin{align}
h & =J_{1}J_{2}\left(1-f\right)\cos\varphi_{2},\label{h_1}
\end{align}
taking the derivative of (\ref{h_1}) with respect to $\varphi_{2}$
to find the maximum value of $h$. Then,
\begin{align*}
\frac{\partial h}{\partial\varphi_{2}} & =-J_{1}J_{2}\left(1-f\right)\text{sin}\varphi_{2}=0,
\end{align*}
which is solved for $\varphi_{2}=0,\,\pi.$ If we put these results
separately into (\ref{h_max_1}) we get 
\begin{align}
h\left(\varphi_{1}=0,\,\varphi_{2}=0\right) & =J_{1}J_{2}\left(1-f\right)<0,\label{h_2}\\
h\left(\varphi_{1}=0,\,\varphi_{2}=\pi\right) & =-J_{1}J_{2}\left(1-f\right)>0.\label{h_3}
\end{align}
Therefore, $h$ is maximum for (\ref{h_3}). The second possibility
is $\varphi_{2}=\frac{\pi}{2}$. Again (\ref{h_max_1}) becomes for
this choice as follows;
\begin{align}
h & =J_{1}J_{2}\sin\varphi_{1},\label{h_4}
\end{align}
note that there is no $f$ dependence. The maximization condition
of (\ref{h_4}) are
\begin{align*}
\frac{\partial h}{\partial\varphi_{1}} & =J_{1}J_{2}\text{cos}\varphi_{1}=0,
\end{align*}
$\varphi_{1}=\frac{\pi}{2},\,\frac{3\pi}{2}$. Putting these into
(\ref{h_max_1}) we get 
\begin{eqnarray}
&& h\left(\varphi_{1}=\frac{\pi}{2},\,\varphi_{2}=\frac{\pi}{2}\right)  =J_{1}J_{2},\label{h_6}\\
&& h\left(\varphi_{1}=\frac{3\pi}{2},\,\varphi_{2}=\frac{\pi}{2}\right) =-J_{1}J_{2}.\label{h_7}
\end{eqnarray}
For this case, $h$ is always a maximum for (\ref{h_6}). Here note
that when $f<2$ (\ref{h_3}) becomes smaller than (\ref{h_6}). Then
for $f<2$ the spins point in the same direction and are perpendicular
to the line joining them. On the other hand, for $f>2$ (\ref{h_3})
is larger than (\ref{h_6}) so the spins are anti-parallel and point along the line joining them. Namely
for massive gravity the spins flip at $m_{g}r\approx1.62$.

\end{document}